\documentclass[preprint,prx,longbibliography]{revtex4-1}
\usepackage{subcaption}
\usepackage{graphicx}
\usepackage{comment}
\usepackage{amsmath,amssymb,amsthm} 
\usepackage{scalerel,stackengine}
\usepackage{bm}
\usepackage{verbatim}
\usepackage{float}

\begin{document}
	\title{ Simulating Physics with Computers}
\newcommand{\orcidauthorA}{0000-0002-0819-698X} 

\author{ S\o ren  Toxvaerd  }

\affiliation{  Department of Science and Environment, Roskilde University, Postbox 260, DK-4000 Roskilde, Denmark}
\email{st@ruc.dk, Submitted to \textit{Symmetry}}

\begin{abstract}
Feynman gave in 1982 a keynote speech \textit{Simulating Physics with Computers} (Int. J. Theor. Phys.  {\bf 21}, 467  (1982)) 
in which he talked  ``...about the possibility...that the computer will do
exactly the same as nature". The motivation was that: ``...the physical world is quantum mechanical, and therefore
the proper problem is the simulation of quantum physics".
Here I try after more than forty years to answer Feynman's question of whether it is possible to perform exact computer simulations.
 Many computer simulations are not exact, they contain mean field approximations that disobey the symmetry
in the quantum dynamics with Newton's third law, e.g. almost all astrophysical simulations of galaxy systems.
After a review of computer simulations and the problems of simulating real systems, I argue that Newton's discrete dynamics, which is
used in almost all computer simulations and which is exact in the
same sense as Newton's analytic dynamics, is the classical limit path 
of Feynman's quantum paths.
 However, the physical world is not known exactly and it is much more complex than any simulated systems, and so far no real systems have been simulated exactly.
Hence, more than forty years later, and after hundreds of thousands of computer simulations of the physical system's dynamics
the answer to Feynman's question is still negative.
But although it is not possible to simulate the dynamics exactly for any real systems, 
the simulations have been and will be of great use in Natural Science.
\end{abstract}

\maketitle

\vspace{2pc}

\section{Introduction}

Richard P. Feynman gave in 1982 a keynote speech \textit{Simulating Physics with Computers} \cite{Feynman1981}
in which he wanted  ``...to talk about the possibility that there is to be an $exact$ simulation, that the computer will do
$exactly$ the same as nature." The motivation was that: ``the physical world is quantum mechanical, and therefore
	the proper problem is \textit{the simulation of quantum physics}".
This article tries after more than forty years,  to answer Feynman's question of whether it is possible to perform an exact computer simulation of the
dynamics of nature.

 Almost all computer simulations of the dynamics of real systems are with algorithms for Newton's classical dynamics.
  Newton's classical $analytic$ dynamics  can, however, only be obtained exactly for a few systems, e.g. the harmonic oscillator, but
	the exact analytic dynamics of a conservative system  of objects with classical dynamics 
	has some characteristics: the dynamics is time reversible, symplectic, and with
	three dynamical invariances  (momentum, angular momentum, and energy). 
 In Feynman's talk he points out that  $exact$
 dynamics is quantum dynamics (QD),  and Feynman also noted that there is a difference between
	\textit{quantizing} and \textit{discretizing}, and that QD is in the coherent space \cite{citef}. However, so are almost all
	computer simulations for classical dynamics. Furthermore, the simulations are not with analytic but discrete dynamics as QD, and where
	the actions of force quants  result in discrete changes of positions in a coherent space.
The formulation of \textit{exact classical discrete dynamics} 
	was solved a long time ago by I. Newton in \textit{Proposition I} in the first part of Newton's
	book  $Principia$ \cite{Newton1687}.
\textit{Proposition I}  is Newton's algorithm for  discrete classical dynamics 
	\cite{Toxvaerd2023}.  In computer simulations, the algorithm appears
	with a variety of names, e.g. the Verlet algorithm \cite{Verlet1967}, or the leap-frog algorithm, and it
	is used in almost all Molecular Dynamics (MD) simulations 
 of classical, as well as semiclassical MD simulations.
	\textit{Proposition I}  is given in the next section, and the algebraic formulation  and
	the proof that Newton's discrete dynamics is exact in the same manner as
	Newton's analytic Classical Mechanics is given in \cite{Toxvaerd2023}.

	Here I will argue that Newton's  discrete dynamics is the classical  discrete limit path in the coherent space of
	Feynman's quantum paths.	
	That it is  a  discrete dynamics, and not an analytic dynamics, which is the limit path of QD was in fact suggested shortly after Feynman's talk by Nobel Laurate T. D. Lee
and coworkers
\cite{Lee1983,Friedberg1983}.

More than forty years after Feynman's talk, the article tries to answer his question based on today's
	knowledge of computer simulations of  real systems.

  \begin{figure}
	  \begin{center}	  
 	 \includegraphics[width=8.6cm,angle=0]{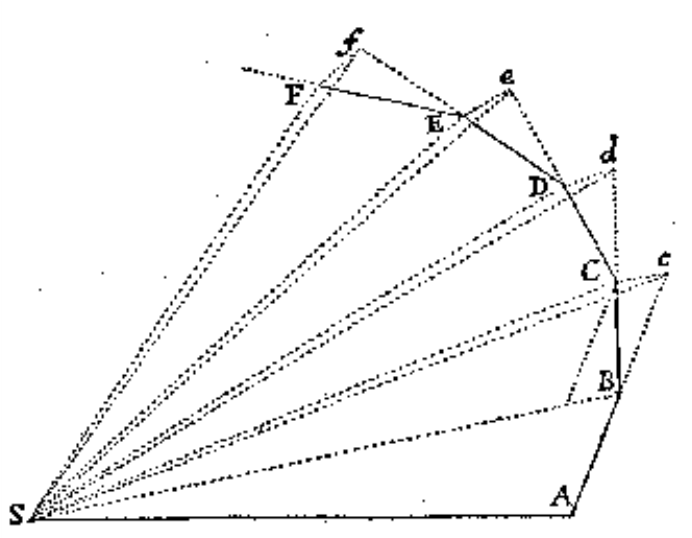}
 	 \caption{  Newton's figure at $Proposition$ $I$ in Principia, with the formulations of the discrete dynamics. 
	  The discrete positions are A: $\bf{r}$$(t-\delta t)$;  B: $\textbf{r}(t)$;  C: $\textbf{r}(t+\delta t)$, etc.. The deviation cC
		  from the straight line ABc (Newton's first law) is caused by a  force from  the position S at time $t$.}
	  \end{center}		
  \end{figure}	   
	\section{Proposition I}\label{sec2}

Newton  begins \textit{Principia} by  \textit{Proposition I}. The English translation is

\textit{Of the Invention of Centripetal Forces.}\\
  PROPOSITION I. Theorem I.\\
 \textit{The areas, which revolving bodies describe by radii drawn to an immovable centre of force do lie in the same immovable planes, and
 are proportional to the times in which they are described}.\\
\textit{ For suppose the time to be divided into equal parts, and in the first part of time let the body by its innate force describe the right line
AB. In the second part of that time, the same would (by Law I.), if not hindered, proceed directly to c, along the line Bc equal to AB; so that the radii
AS, BS, cS, drawn to the centre, the equal areas ASB, BSc, would be described. But when the body is arrived at B,
\textbf{suppose that a centripetal force acts at once
with a great impulse}, and, turning aside the body from the right line Bc, compels it afterwards
to continue its motion  along the right line BC. Draw cC parallel
to BS meeting BC in C; and at the end of  the second part of the time, the body (by Cor. I of Laws) will be found in C, in the same plane with the
triangle ASB. Join SC, and,  because SB and Cs are parallel, the triangle SBC will be equal to the triangle SBc, and therefore also to the
triangle SAB. By the like argument, if the centripetal force acts successively in C, D, E, \& c., and makes the body,
in each single particle of time, to describe the right lines CD, DE, EF, \& c., they will all lie in the same plane;
and the triangle SCD will be equal to the triangle SBC, and SDE to SCD, and SEF to SDE. And therefore, in equal times, equal areas are described in on immovable plane:
and, by composition, any sums SADS, SAFS, of those areas, are one to the other as the times in which they are described. Now let the number of
those triangles be augmented; and their breadth diminished in infinitum; and (by Cor. 4, Lem III) their ultimate perimeter ADF will be a curve line:
and therefore the centripetal force, by which the body is perpetually drawn back from the tangent of this curve, will act continually; and any described
areas SADS, SAFS, which are always proportional to the times of description, will, in this case also, be proportional to those times.} Q. E. D.

Newton illustrated \textit{Proposition I} by a figure (Fig. 1).
The assumption ..\textit{\textbf{suppose that a centripetal force acts at once with a great impulse}}..
  is highlighted here. It is a central part in  the proposition. The force at times between a discrete time step is zero,
 time and forces are \textit{quantized}, and
 the  change in  positions in a coherent space is \textit{discretized} proportional to the actions of the forces.

 According to $\textit{Proposition I}$, a new  position in Newton's  discrete dynamics, $\textbf{r}_i(t+\delta t)$, at time $t+\delta t$ of an object  
$i$ with the mass $m_i$   is determined by
the force quant $\textbf{f}_i(t)$ acting on the object   at the discrete position $\textbf{r}_i(t)$  at time $t$. The quant  causes a discrete change
of  the direction,  and it  changes the momenta and the positions  $ \delta \textbf{r}_i$ proportional to $\delta t$ with  $ \delta t \textbf{f}_i(t)= \overrightarrow{cC}$ (see Figure 1).
The positions inbetween the force quants are changed from  $\overrightarrow{AB}= \textbf{r}_i(t)-\textbf{r}_i(t-\delta t)=\overrightarrow{Bc}$
to  $\overrightarrow{BC}= \textbf{r}_i(t+\delta t)-\textbf{r}_i(t)$, i.e
\begin{equation}
	\overrightarrow{BC}=\overrightarrow{AB}+ \overrightarrow{cC},
 \end{equation}
or
\begin{equation}
	 m_i\frac{\textbf{r}_i(t+\delta t)-\textbf{r}_i(t)}{\delta t}
			=m_i\frac{\textbf{r}_i(t)-\textbf{r}_i(t-\delta t)}{\delta t} +\delta t \textbf{f}_i(t).	
 \end{equation}
 The algorithm, Eq. (2), is usual presented  as the ``the Verlet algorithm"
 \begin{equation}
	 \textbf{r}_i(t+\delta t)=2\textbf{r}_i(t)-\textbf{r}_i(t-\delta t) +\delta t^2/m_i \textbf{f}_i(t),	
 \end{equation}
 or the  ``leap frog" algorithm for the velocities
\begin{eqnarray}
\textbf{v}_i(t+\delta t/2)=  \textbf{v}_i(t-\delta t/2)+ \delta t/m_i  \textbf{f}_i(t) \nonumber \\
\textbf{r}_i(t+\delta t)=\textbf{r}_i(t)+\delta t\textbf{v}_i(t+\delta t/2).
\end{eqnarray}

 Newton's discrete dynamics, Eq. (3) or   Eq. (4,)
	is used in almost all  simulations 
 of classical, as well as semiclassical  simulations.
The trajectory A,B,C,D,.. 
in  Figure 1 in \textit{Proposition I} is a discrete classical QD path
but the difference between analytic- and discrete dynamics is 
 proportional to the square of $\delta t=$ Planck time ($\approx 10^{-44}$ sec) and negligible \cite{Toxvaerd2014}.

 \section{Simulating  with Computers}\label{sec3}

The title of Feynman's talk is \textit{Simulating Physics with Computers}, but today computer simulations are used not only in physics but in almost all subdisciplines
 in Natural Science. The first section after Feynman's Introduction is \textit{ 2. Simulating Time}, and thus he limited this talk to cover the
 part of computer simulations that concerns the simulations of the time evolutions of physical objects. These
 simulations are often named Molecular Dynamics, with a misleading name since they cover all kinds of simulations with objects from atoms 
 to galaxies. Feynman excludes the many
 statistical mechanics simulations, which often are named Monte Carlo simulations (MC)  with a name that refers to the statistical probability of an event in  MC.

 It appears from his speech that Feynman has a supercomputer in mind and with an enormous hardware capability, and these computers exist today and
 might be extended to quantum computers in the future (which probably will bring humanity to a new era in the ``artificial intelligence" world).
 But computer simulations are not hardware, it is software and for the dynamic evolution of a physical system, they are algorithms. Newton was the first to formulate
 an algorithm in \textit{Proposition I} and he (and his students) used it to calculate the future of the Solar system. Newton lived in the rational era and
 the evolution of the Solar system was believed to be  ``exactly" predicted by Newton's ``simulations", a fact which revolutionized the philosophy. 
The first MD simulations appeared  in the sixties
\cite{Rahman1964,Verlet1967}, almost three hundred years after $Proposition$'s publication 
 and today there have appeared at least many hundred thousand articles with MD,  or might be more than 
 a million.

 An MD simulation is usually
 for a system with $N \approx 1000-10000$ objects, with periodical boundaries to simulate a continuum,
 with approximated force fields, with  truncated force fields at long-range interactions \cite{Toxvaerd2011}, and with discrete time  increments
 $\delta t \approx 10^{-14}$ sec much larger than Planck time. In astrophysics, the simulations are
 for a much larger system, but with mean-field approximations for long-range interactions. These approximations disobey fundamental physical laws
 and destroy the exactness of the discrete classical dynamics (see later).
Many simulations are often performed using packages with ``black box" software to
 the simulations. In general, the simulations are not exact since they appear with a series of approximations of the models of the real systems.

 The many computer simulations have, however, changed our insight into the dynamic evolution of real systems, and 
the success of MD was acknowledged by a Nobel Prize in 2013 \cite{Nobel2013}.

\section{ Newton's third law, Feynman's quantum paths and QD entanglements}\label{sec4}

Feynman's talk was about simulating QD  in \textit{the physical world} and  
 several properties must be clarified before one can answer Feynman's question of whether it is possible to perform an
 exact computer simulation of a real system, e.g.:

A.  Is Newton's third law valid in QD?

B.  Is gravity  quantized, and with gravitational waves? 

C. Does the interaction between baryonic object pairs with gravitation follow QD, with Feynman's quantum paths and QD entanglement?

All three  questions have been heavily debated in the past.

\subsection{Newton's third law}
Newton began $Principia$ by postulating his three laws, and he
justified the third law from daily life experience in the subsequent $Scholium$ in $Principia$. Newton's law  of the inverse square gravitational attraction (ISL), which entangles
pairs of baryonic objects according to his third law is, however, formulated much later in $Principia$ on page 401 in the Third Book \textit{Mundi Systemate}, and
where Newton first formulated four philosophic principles for Natural Science. The first and in an English translation reads:\\
\textit{ We are to admit no more causes of natural things than such as are both true and sufficient to explain their appearances.}\\
This principle might not be fulfilled by assuming the existence of dark matter in the Universe (see later).

The validity of the third law has been debated right up to our days, for a historical review see \cite{Domski2022}. The symmetry of the interactions
ensures the conservations of momentum and angular momentum for a conservative system (Noerthe's theorem \cite{Noerthe1918}). 
It has been questioned  whether 
Newton's third law is valid for  conservative electromagnetic systems. It was doubted by P. Cornille  in his review of the applications of
Newton's third law in physics \cite{Cornille1999}. According to Cornille, the law is not valid for electromagnetic systems,
and Cornille claims that it can be disproved for a charged
conductor at rest which is predicted to set itself in motion and accelerate its center of mass or rotate without
external help. It is a breach of the second law of thermodynamics,
but it is a quality which would be of great technical importance. However, no such conductors have been reported and a recent article
confirms the validity of Newton's third law for these electromagnetic systems \cite{Epp2023}.

Newton's third law 
 ensures the conservation of momentum and angular momentum  for a conservative system \cite{Noerthe1918}.
 However, many computer simulations e.g. simulations of galaxies, use mean field approximations which   disobey
Newton's third law.  Simulations of galaxies with the ensemble of ``suns" with gravitational forces have been performed for many decades.  The dynamics of the galaxies are obtained
 by the  Particle-Particle/Particle-Mess (PPPM)  method  \cite{Hockney1974,Klypin1983,Centrella1983}, which is
 a mean-field approximation of the long-range forces in the system where 
 a mass unit is moving in \textit{the collective field of these forces}, and the  Poison equation
 for the PPPM grid is solved numerically.
The galaxies are believed to be unstable unless they are stabilized by dark matter and the indirect proof of its existence is in fact 
based on extensive computer simulations \cite{Springel2005,Schaye2010,Dubois2014,Vogelsberger2014,Dubois2016,Price2018,Weinberger2020,Ludlow2021}.
Cosmological simulations of galaxies are reviewed in \cite{Vogelsberger2020}.
The justification for assuming the existence of dark matter is reviewed in \cite{Bahcall1995,Bahcall2015}. The justification is not only
 based on the simulations of galaxy systems but  on  several experimental observations of 
 galaxies.  The mass distribution and the rotational
 velocities  are non-Newtoinian/Kepler-like with a substantial baryonic matter in the halos of the galaxies and with
 a higher rotational velocity than in a system with classical dynamics. These facts  have supported the hypothesis of the existence of dark matter.

 Also the alternative MOND formulation \cite{Milgrom1983} 
 for classical celestial dynamics disobeys Newton's  third law \cite{Felten1984,Toxvaerd2024}, and
 computer simulations of galaxies reveal that the lack of conserved angular momentum destabilizes
 the galaxy systems \cite{Toxvaerd2024}. 
But the recent exact  simulations of models of galaxy systems \cite{Toxvaerd2024} 
 demonstrate that the stability and  a correct rotation profile of the galaxies can be obtained purely from  classical dynamics that obey Newton's laws
 by modifing Newton's law of universal inverse square attraction.

\textit{ In conclusion}: So far no experimental evidence exists that Newton's third law is not valid for any systems, including the dynamics of
 galaxy systems and quantum electromagnetic dynamics (QED). Concerning computer simulations, the simulations with mean field attractions or MOND 
disobey Newton's third law  and they are not exact.

\subsection{ Gravity quantization and gavitational waves}

Feynman assumed that the dynamics appear with positions in a coherent space, and so do the computer-generated dynamics, but with discrete changes of the positions.
Lee and Friedberg suggested quantization of changes in positions \cite{Lee1987,Friedberg1994}, and quantization of the gravitational
attraction has been suggested for a long time ago to overcome several problems 
with a continuum formulation of general relativity  \cite{Regge1961}. Since then a series of articles with
theories of quantum gravity based on the path integral approach of Feynman have appeared. For reviews see \cite{Ambjorn2012,Loll2020,Ambjorn2021}.

Stephen Hawking wrote in 1999: \textit{So what the singularity theorems are really telling us, is that the universe had a quantum origin,
and that we need a theory of quantum cosmology, if we are to predict the present state of the universe} \cite{Wiki}.
Quantum cosmology is now a discipline in astrophysics with countless articles \cite{QuantumCosmos2022}.

Einstein predicted the existence of gravitational waves caused by
acceleration of mass units
in the Einstein-de Sitter Universe.
For a review of gravitational waves see \cite{Cota2016}.  One of 
the greatest success of the Hubble telescope was the experimental detection of gravity waves \cite{Abbott2016,Abbott2016a},
and the  accelerated expansion of the Universe \cite{Riess1998,Perlmutter1999}.
The equal behavior of electromagnetic and gravitational waves was the experimental verification of gravity quantization.
The Universe expands with the relative Hubble expansion, which is an accelerating expansion with an intensive expansion at every point
in our three-dimensional space (3D).  The accelerated expansion of
the space with baryonic units will create and propagate gravitational waves with
an inverse square intensity from the baryonic units in a 3D space. This fact explains the common behavior of gravitation and the Coulomb ISL attraction.
The gravitation maintains a long-ranged ISL attraction due to the symmetry break at Hawking's 
QD origin of the Universe with the lack of antimatter.

Newton's discrete algorithm, which is a central difference algorithm has been extended to include the central difference effect
on positions due to the Hubble expansion  \cite{Toxvaerd2022a}, and the extended algorithm is 
used to simulate models of galaxies \cite{Toxvaerd2022a,Toxvaerd2024}.
 
\textit{ In conclusion}:  There is no conflict  in assuming quantum gravity  and  space-time quantization in connection with  Feynman's  path integral formulation
of QD.\\

\subsection{ Barionic object pairs with gravitation QD, with Feynman's quantum paths and QD entanglements }

Newton's third law ensures the QD entanglement between two baryonic units $i$ and $j$, and if Newton's law of gravitational attraction is caused by
gravitational waves which, with the speed of light moves from $i$ and $j$, respectively, then the attraction on $i$ and $j$ appears synchronously  and ensures the
entanglement between $i$ and $j$. 
However, although gravitation might be quantized and with QD entanglement, Newton's universal gravitational attraction is not included in the standard model. This fact
has given rise to a series of proposed modifications of Newton's inverse square law (ISL) for gravitational attraction to an inverse attraction (IA) for longer distances
with the intention of unifying gravity with particle physics, so far without success  \cite{Fischbach2001,Adelberger2003,Lee2020,Henrichs2021}. 

Einstein also proposed a modification of the ISL law \cite{Einstein1928}, 
and series of modified  gravity  in cosmology including IA attractions have been proposed, for a review see \cite{Capozziello2011,Clifton2012}.
The ISL might also be modified by gravitational lensing \cite{Renn1997,Toxvaerd2024}. 
The  cosmological effect of  gravitation
must be very long-range in favor of e.g. an IA attraction since galaxies are known to appear in clusters with bound rotations. The Milky Way and Andromeda galaxies belong to the \textit{Local Group} \cite{Redd2018},
and these galaxy clusters are again collected in the \textit{Virgo super collection of galaxies clusters} \cite{Mei2009}. 

\textit{ In conclusion}: Although there is nothing exclusive in assuming quantum gravity, the gravitational attraction is not included
in the standard model and there is a need to explain the stability, mass distribution, and rotational velocity of galaxies either by including dark matter in the universe
or by modifying Newton's ISL, e.g. to an IA long-range attraction.

\section{Reply to Feynman's question}

When Feynman in his talk ``wants to talk about the possibility that there is to be an $exact$ simulation, that the computer will do
$exactly$ the same as nature", there is a fundamental problem because we do not exactly know what nature does.  
Feynman lists a series of problems concerning performing exact computer simulations, and no doubt his answer to the question was negative, it is not
possible. Nevertheless, hundreds of thousands of computer simulations have been published since his talk,
and they have led to a much better understanding of the dynamics of real systems. However, almost all
computer simulations contain a series of approximations, e.g., mean field approximations and approximations of the range of the forces. 
 They are not strictly exact, but in the best case, they are qualitatively correct.

Feynman pointed out  that: ``...the physical world is quantum mechanical, and therefore
the proper problem is the simulation of quantum physics".
Newton formulated in $Principia$ his three laws, the ISL  law for gravitation, and found the analytic solution for Kepler's equation for pairs of objects with the ISL.
The correct dynamics is, however, QD as Feynman pointed out, and if so then Newton's analytic dynamics is not the classical limit path in QD.
It is Newton's discrete
dynamics, formulated in \textit{Proposition I} at the beginning of $Principia$ which is the classical limit path of Feynman's quantum paths,
and with the QD entanglement given by Newton's third law. The difference between the discrete and the analytic classical dynamics is, however, marginal, and furthermore, there exists a ``shadow  Hamiltonian"
\cite{Toxvaerd1994} where the discrete positions are located on the analytic trajectories for the analytic dynamics for the shadow  Hamiltonian.
So there is no qualitative difference between Newton's analytic and discrete classical dynamics \cite{Toxvaerd2023}.
Newton lived  1643-1727 in the \textit{Age of Enlightenment}, and he believed that his analytic dynamics were God's exact dynamics.

 Nature consists of elementary particles, atoms, molecules,..., planets, solar systems,  galaxies, and perhaps black matter, 
and with different forces where not all are given by the standard model.
All the computer simulations for  real systems  are with  drastical simplifications of the systems. The simulations of models of dwarf galaxies in \cite{Toxvaerd2024}are
claimed to be exact simulations. But they are only exact in the sense that they are without any approximation of the classical Feynman path of QD and they
are still a drastic simplification of even the smallest dwarf galaxy \cite{Simon2019}.
Hence, more than forty years after Feynman formulated his question the answer is still negative, the computer simulations are not exact.
But in the real world,
simulations with  Newton's discrete algorithm have been and will be of great use in natural science.

\section*{Acknowledgements}
Niccol\~{o} Guicciardini  is gratefully acknowledged.

\end{document}